# A two-state model of twisted intramolecular charge-transfer in monomethine dyes

*Seth Olsen* and *Ross H. McKenzie*

School of Mathematics and Physics, The University of Queensland, Brisbane QLD 4072 Australia

*seth.olsen@uq.edu.au

**Abstract** A two-state model Hamiltonian is proposed to model the coupling of twisting displacements to charge-transfer behavior in the ground and excited states of a general monomethine dye molecule. This coupling may be relevant to the molecular mechanism of environment-dependent fluorescence yield enhancement. The model is parameterized against quantum chemical calculations on different protonation states of the green fluorescent protein chromophore (GFP), which are chosen to sample different regimes of detuning from the cyanine (resonant) limit. The model provides a simple yet realistic description of the charge transfer character along two possible excited state twisting channels associated with the methine bridge.

It describes qualitatively different behavior in three regions that can be classified by their relationship to the resonant (cyanine) limit. The regimes differ by the presence or absence of twist-dependent polarization reversal and the occurrence of conical intersections. We find that selective biasing of one twisting channel over another by an applied diabatic biasing potential can only be achieved in a finite range of parameters near the cyanine limit.

1. **Introduction**

Excited-state twisting processes in monomethine dyes (di- and triarylmethanes, monomethine cyanines) are of technological interest because their suppression leads to environment-dependent fluorescence quantum yields[1-3]. In several cases, the steady-state fluorescence quantum yield increases by four orders of magnitude or more upon binding to biomolecules[1, 4-8]. Examples include Malachite Green[6, 9], Crystal Violet[6], Oxazole Yellow[9], and various derivatives of the green fluorescent protein (GFP[10-12]) chromophore[1, 7, 8, 13]. Dyes that display this behavior are useful as biotechnological stains and sensors[10, 12, 14]. In some cases, the fluorescence in the bound state can be switched on and off by a photoconversion process[15-17]. This behavior has proven useful in high-resolution microscopy[18, 19]. Interest has focused on the molecular origin of the fluorescence enhancement, so as to inform the design of better fluorogens[2]. This raises a challenge to develop simple, general models of twisting behavior that can be used to model and understand the physics of fluorescence enhancement in dissipative and complex interacting environments.

Excited-state twisting processes have been invoked to explain the sub-picosecond fluorescence decay times and 1-10ps ground state recovery times that are observed for monomethine dyes in low-viscosity solutions at ambient pressure and temperature[20-30]. Twisting of the methine bonds in the excited state is predicted to lower the potential energy of the excited state, and to lead to charge-localized (for neutral dyes, charge-separated) states from which internal conversion is possible due to a reduced adiabatic energy gap[31, 32]. In several cases, electronic structure computations[33-38] have predicted the existence of low-energy conical intersection seams with significant twist, which would be expected to facilitate ultrafast deactivation. The conical intersections themselves arise from the crossing of diabatic states with distinct charge localization, and simulations indicate that coupling of solvation behavior to the charge-transfer behavior can significantly affect the rate of decay[35].

The excited-state twisting process in monomethine systems has been invoked as an example of an environment-controlled process with no intrinsic barrier[21, 28, 39-44]. The dynamics have been described with theoretical models invoking overdamped motion[40-45]. Although the coupling of charge-transfer behavior to twisting displacements has been predicted for some time[26], the theoretical models usually do not explicitly describe this. Contributions to the twisting rate from electrostatic solvent reorganization have not been addressed. Experiments on several common dyes suggest that solvent viscosity and polarity both affect the rate, although the influence of the former dominates the dynamics in solution[21, 46]. Other experiments suggest that balance of viscosity and polarity effects may depend on the chemical identity of the dye used[20, 25, 47].

There are not one, but two accessible twisting channels in the excited state for dyes that are near the cyanine (resonant) limit. The ground state of such dies is characterized by an equivocal superposition of bonding states, while the excited state by an superposition of diradical character for both bonds of the bridge. Twisting in the excited state can occur about both bonds, but the different channels are distinguishable by coupling to distinct charge-transfer polarizations. This behavior is predicted by very simple and general theoretical considerations[31, 32]. It has not received much attention in the discussions of spectroscopic experiments, wherein it is often assumed that the dynamics can be described by a single effective coordinate[40-43, 45].

In order to identify whether twist-dependent charge transfer behavior and pathway bifurcation effects are relevant to understanding the twisting dynamics of monomethines, it is necessary to formulate simple physical models that can describe these phenomena for a general case. *This is the point of this paper*.

The green fluorescent protein (GFP[14]) chromophore is an interesting case for probing the effects of charge-transfer and twisting pathway bifurcation on the dynamics, because protonation of different ends of the oxonol chain allows sampling from electronic structures both near and far from the cyanine limit[48, 49]. The effects of protonation on the twisting pathways are qualitative; protonation at the phenolic oxygen is apparently sufficient to effectively remove one of two possible twisting channels, and also eliminates the twist-dependence of the excited-state charge localization[38, 50].

Experimentally, the viscosity dependence of the decay of synthetic GFP chromophore models is weak[20, 51, 52]. It has been remarked that the fluorescence decay times of fluorescent protein chromophore models in different protonation states can vary by up to an order of magnitude[53]. This effect cannot be accounted for by a purely viscosity-dependent mechanism (as has been invoked in other systems[40, 42, 45]), because titration should not significantly change the molecular volume nor the surface area[22]. A study of a wide range of different chemical derivatives indicated that several mechanisms may be active, including internal conversion by large-amplitude motion but also possibly hydrogen-bond-assisted decay and/or charge-transfer, depending upon the derivative and the solvent used[25, 54]. These experiments therefore highlight the pressing need for models that can describe changes in the potential surface and the charge-transfer behavior at different proximities to the cyanine limit.

The excited-state twisting processes of the chromophores of GFP variants are of direct technological interest because of their implication in medium-dependent emission properties[55], reversible photoconversion processes[56-59] and light-activated assembly of split-protein constructs[60, 61].

In this paper, we describe a simple two-state model Hamiltonian that can describe the potential surface and the charge-transfer dependence of monomethine dyes at different detuning from the cyanine limit. The model is a generalized Mulliken-Hush electron transfer model[62, 63], wherein the diabatic state energies and electron transfer matrix element are coupled to twisting displacements about the methine bridge bonds. The model describes the bifurcation of twisting channels, as well as twist-dependent polarization, in dyes near the cyanine limit. It also describes the elimination of twisting channels, and the loss of twist-dependence of the polarization, in dyes far from the limit. As a demonstration of the capacity of the model, we parameterize it against a selection of protonation states of the green fluorescent protein chromophore that sample electronic structures both near and far from the cyanine limit[48, 49]. We claim that the model invokes a level of specific chemical detail intermediate between those used to generally describe barrierless photoisomerization in monomethines[22, 40, 42, 64] and the atomistically (or electronically) detailed simulations that have indicated strong effects related to the development of charge transfer behavior[35]. It is in the spirit of recent "essential state" models that have successfully applied to model the spectra of organic chromophores[65-67].

The paper proceeds as follows. Section 2 describes the model Hamiltonian and the basic theoretical analytical framework, as well as providing specific details of the calculations used to parameterize it for different oxonol protonation states of the GFP chromophore (c.f. Scheme 1). In section 3, we outline several important results. We show that the model reproduces the essential features of the quantum chemical calculations with respect to the occurrence of low-energy twisting channels and their coupling to the charge transfer behavior. We show that there is a connection between the occurrence twist-dependent charge-transfer polarization reversal and the occurrence of conical intersections in the model. The model describes qualitatively different regimes of electronic structure where the driving forces for the distinct twisting channels may or may not be tuned by the application of a diabatic biasing potential (e.g. by protonation or an external field). Twisting channel selection can only be accomplished in a limited region of parameter space around the cyanine limit; this region also limits the range for which twist-dependent polarization reversal and twisted conical intersections may occur. We provide explicit formulas for the driving forces along the distinct channels, and their dependence on the diabatic biasing potential. Section 4 discusses the results in the context of recent experimental and theoretical studies. Section 5 concludes the paper.

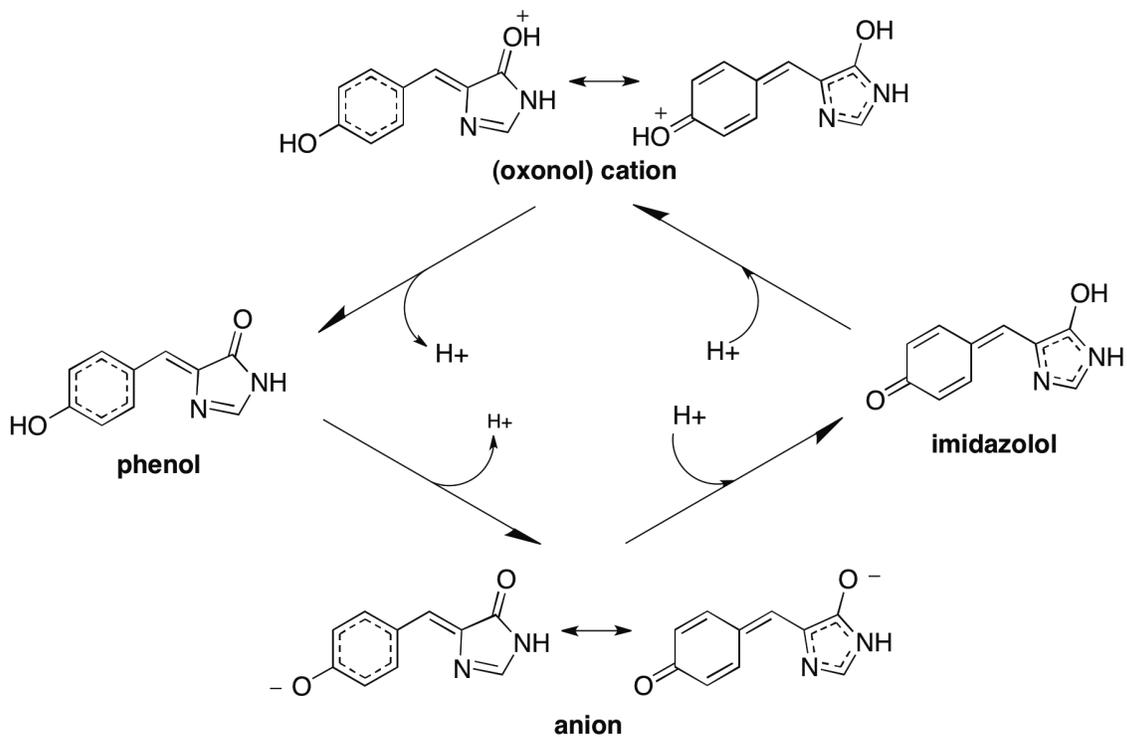

SCHEME 1

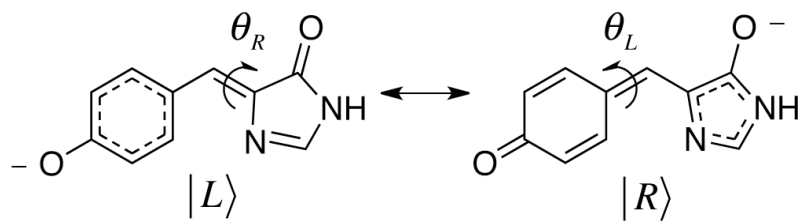

SCHEME 2

## 2. Theoretical Model and Parameterization

### 2.1 Model Hamiltonian

The model Hamiltonian is defined on a basis of diabatic[68] states $|L\rangle$ and $|R\rangle$ (c.f. Scheme 2) and is parameterized as

$$\langle L|H(\theta_L,\theta_R)|L\rangle = \frac{1}{2}\delta + 2\gamma_R \sin^2\theta_R + 2\gamma'_L \sin^2\theta_L$$

$$\langle R|H(\theta_L,\theta_R)|R\rangle = -\frac{1}{2}\delta + 2\gamma_L \sin^2\theta_L + 2\gamma'_R \sin^2\theta_R \qquad (2.1)$$

$$\langle L|H(\theta_L,\theta_R)|R\rangle = \frac{1}{2}\varepsilon \cos\theta_L \cos\theta_R$$

The action of the Hamiltonian on the electronic state basis is dependent on the values of the "dynamical variables" $\theta_L$ and $\theta_R$, which represent the state of twist of the methine bridge bonds (c.f. Scheme 2).

The functional form of the dependence on $\theta_L$ and $\theta_R$ can be justified using valence-bond models of chemical bonding[69-72]. The $\sin^2\theta$ terms on the diagonal represent the energetic cost associated with twisting the bonds in the different diabatic states. Based on the diagrams in Scheme 2, one would suppose that the energy associated with twisting the double bond is larger than the single bond, so that $\gamma_i > \gamma'_i$. In the next section, we will show that this is the case. The $\cos\theta_L \cos\theta_R$ dependence of the coupling element, on the other hand, can be intuitively justified as the product of the overlaps of $\pi$ orbitals on the each of the rings with the bridge. These functional forms can be derived by writing down a Hückel-Hubbard model Hamiltonian[73] for a system of three electrons in three $\pi$ orbitals – with the orbital character encoded in a $\cos\theta_i$ dependence of the hopping elements on the bridge bonds – and performing a second order canonical (Van Vleck) transformation[74] to isolate the two-dimensional diabatic states suggested by Scheme 2. This general approach has been used to derive effective valence-bond Hamiltonians for describing electronic structure calculations[70-72].

The adiabatic states and their energies are completely specified by a dimensionless parameter $\lambda(\theta_L,\theta_R)$, which is a function of the twist angles.

$$\lambda(\theta_L,\theta_R) \equiv \frac{\delta + 2(\gamma_R - \gamma'_R)\sin^2\theta_R - 2(\gamma_L - \gamma'_L)\sin^2\theta_L}{\varepsilon \cos\theta_L \cos\theta_R} = \cot 2\phi \quad (2.2)$$

Here $\phi(\theta_L,\theta_R)$ is the mixing angle defining the transformation to the adiabatic representation. The adiabatic states are then

$$\begin{aligned}|0\rangle &= \cos\phi(\theta_L,\theta_R)|L\rangle + \sin\phi(\theta_L,\theta_R)|R\rangle \\ |1\rangle &= -\sin\phi(\theta_L,\theta_R)|L\rangle + \cos\phi(\theta_L,\theta_R)|R\rangle\end{aligned} \quad (2.3)$$

The adiabatic state energies are

$$\begin{aligned}E_0 &\equiv \langle 0|H(\theta_L,\theta_R)|0\rangle = \bar{E}(\theta_L,\theta_R) - \frac{1}{2}\Delta E(\theta_L,\theta_R) \\ E_1 &\equiv \langle 1|H(\theta_L,\theta_R)|1\rangle = \bar{E}(\theta_L,\theta_R) + \frac{1}{2}\Delta E(\theta_L,\theta_R)\end{aligned} \quad (2.4)$$

where

$$\bar{E}(\theta_L,\theta_R) = (\gamma_R + \gamma'_R)\sin^2\theta_R + (\gamma_L + \gamma'_L)\sin^2\theta_L \quad (2.5)$$

is the state-averaged energy and

$$\Delta E(\theta_L,\theta_R) = \sqrt{(\varepsilon\cos\theta_L\cos\theta_R)^2 + (\delta + 2(\gamma_R-\gamma'_R)\sin^2\theta_R - 2(\gamma_L-\gamma'_L)\sin^2\theta_L)^2} \quad (2.6)$$

is the adiabatic energy gap.

The diabatic representation relevant to (2.1) is taken to be the Generalized Mulliken-Hush (GMH) representation[62, 63], so that the dipole operator projected along the direction of charge-resonance/charge-transfer is diagonal in the diabatic representation[62, 63]. We can then express the dimensionless parameter $\lambda(\theta_L,\theta_R)$ in terms of the adiabatic difference dipole $\Delta\mu(\theta_L,\theta_R)$ and the adiabatic transition dipole $\mu_{01}(\theta_L,\theta_R)$ (projected along the direction of charge-resonance/charge-transfer) as

$$\lambda(\theta_L,\theta_R) = \frac{\Delta\mu(\theta_L,\theta_R)}{\mu_{01}(\theta_L,\theta_R)}$$
$$= \frac{\Delta\tilde{\mu}(\theta_L,\theta_R)}{\sqrt{1-\left(\Delta\tilde{\mu}(\theta_L,\theta_R)\right)^2}} \qquad (2.7)$$
$$= \frac{1}{\sqrt{1-\left(2\tilde{\mu}_{01}(\theta_L,\theta_R)\right)^2}}$$

where we have defined the dimensionless adiabatic difference dipole

$$\Delta\tilde{\mu}(\theta_L,\theta_R) \equiv \frac{\Delta\mu(\theta_L,\theta_R)}{eR_{DA}(\theta_L,\theta_R)} = \frac{\lambda(\theta_L,\theta_R)}{\sqrt{1+\lambda^2(\theta_L,\theta_R)}}, \qquad (2.8)$$

the dimensionless adiabatic transition dipole moment

$$2\tilde{\mu}_{01}(\theta_L,\theta_R) \equiv \frac{2\mu_{01}(\theta_L,\theta_R)}{eR_{DA}(\theta_L,\theta_R)} = \frac{1}{\sqrt{1+\lambda^2(\theta_L,\theta_R)}}, \qquad (2.9)$$

and the diabatic difference dipole moment

$$eR_{DA}(\theta_L,\theta_R) \equiv \sqrt{\left(\Delta\mu(\theta_L,\theta_R)\right)^2 + \left(2\mu_{01}(\theta_L,\theta_R)\right)^2} . \qquad (2.10)$$

We have expressed the diabatic difference dipole as a product of the elementary charge $e$ and an effective charge-resonance/charge-transfer distance $R_{DA}$.

There is a sign ambiguity associated with the square root in (2.10). To resolve the ambiguity in such a way that the dipole observables change continuously with respect to the torsion, it is useful to adopt the convention that $\text{sgn}(eR_{DA}) = \text{sgn}(\mu_{01})$, so that $\text{sgn}(\Delta\tilde{\mu}) = \text{sgn}(\Delta\mu)$.

The model was parameterized against the ground state minimum and two excited-state twisted geometries for four protonation states of the green fluorescent protein (GFP) chromophore (c.f. Scheme 1). The computational protocol used to produce the models is described in more detail in the next section. The six parameters defining the Hamiltonian (2.1) were determined using the adiabatic energies and dipole observables at three geometries corresponding to $(\theta_L, \theta_R) \in \left\{ (0,0), \left(0, \frac{\pi}{2}\right), \left(\frac{\pi}{2}, 0\right) \right\}$ using the relations

$$\delta = \Delta E(0,0) \Delta \tilde{\mu}(0,0)$$
$$\varepsilon = \Delta E(0,0) 2\tilde{\mu}_{01}(0,0) = \Delta E(0,0)\sqrt{1-\left(\Delta\tilde{\mu}(0,0)\right)^2}$$
(2.11)

$$2\gamma_R = \bar{E}\left(0,\frac{\pi}{2}\right) + \frac{1}{2}\left(\Delta E\left(0,\frac{\pi}{2}\right)\Delta\tilde{\mu}\left(0,\frac{\pi}{2}\right) - \delta\right)$$
$$2\gamma'_R = \bar{E}\left(0,\frac{\pi}{2}\right) - \frac{1}{2}\left(\Delta E\left(0,\frac{\pi}{2}\right)\Delta\tilde{\mu}\left(0,\frac{\pi}{2}\right) - \delta\right)$$
(2.12)

$$2\gamma_L = \bar{E}\left(\frac{\pi}{2},0\right) - \frac{1}{2}\left(\Delta E\left(\frac{\pi}{2},0\right)\Delta\tilde{\mu}\left(\frac{\pi}{2},0\right) - \delta\right)$$
$$2\gamma'_L = \bar{E}\left(\frac{\pi}{2},0\right) + \frac{1}{2}\left(\Delta E\left(\frac{\pi}{2},0\right)\Delta\tilde{\mu}\left(\frac{\pi}{2},0\right) - \delta\right)$$
(2.13)

**2.2. Computational Quantum Chemistry**

For each of the protonation states of the GFP chromophore model shown in Scheme 1, we obtained three geometries by optimization on the ground or excited state surface using a $2^{nd}$ order multireference Raleigh-Schrödinger perturbation theory (RS2) model[75] based on a two-state-averaged four-electron, three-orbital complete active space self consistent field[75] (SA2-CAS(4,3)) reference and a cc-pvdz basis set[76] in MOLPRO[77].

The SA2-CAS(4,3) electronic structure model that we use here has been previously applied to the electronic structure of GFP chromophore models[48, 49]. The active spaces are shown in Figure 1, using the Boys(-Edminston-Ruedenberg) orbital representation[78, 79]. The active space structure is analogous to an allylic π-electron system. Natural orbitals and occupation numbers characterizing the active space solutions are also described in Tables S1-S12 of the Supplement. A four-electron three-orbital system is the *minimal* system that can describe the resonant transfer of a singlet bond and an electron pair (c.f. Scheme 1). Olsen has shown that this level of description provides a robust description of the low-energy spectra of diarylmethanes[80]. The dimension of the active space is consistent with simple analyses of the electronic structure of GFP chromophore models that have been presented by other groups[81]. A comparison against calculations with larger bases and active spaces indicates that this model is adequate to describe the relevant physics of the chromophore[82-84]. The excited state of the anionic GFP chromophore model is predicted to be autoionizing in the gas phase[84], but this behavior is suppressed by solvation[85] or in protein[86]. Our ultimate goal is to design models for the description of condensed phase behavior; we are not concerned with the gas-phase autoionization.

We obtained models of the excited-state twisting channels by linear interpolation (in internal coordinates) between the ground state minima and two twisted excited state structures for each form. The coordinates used for the interpolation are described in Tables S13-S25 of the Supplement. This represents a crude model of the twisting channels, because it does not describe relaxation on the excited state surface along degrees of freedom orthogonal to the twisting modes – for example, the bridge bond length alternation modes. These modes are known to play an important role in the photophysics of conjugated dyes generally[65, 66, 87-90] and in fluorescent protein chromophores particularly[91-94]. The relevance of bridge stretching modes to the development of charge-transfer behavior is implicit in Scheme 2. The omission of these modes is in keeping with our current narrow goal of achieving a *simple* model of the coupling between twisting and charge-transfer behavior; strategies for the incorporation of other important modes are deferred for later work.

The data obtained at the geometries above were used to parameterize the model Hamiltonian (2.1) for each case using the relations (2.11) – (2.13). Relevant details of each geometry and optimization include:

1. $(\theta_L, \theta_R) = (0,0)$ : The geometry is planar about both bonds. The geometry was optimized on the ground state surface. No constraints were employed in any case.

2. $(\theta_L, \theta_R) = \left(0, \frac{\pi}{2}\right)$ : The geometry is twisted about the imidazoloxy-bridge bond, while the phenoxy-bridge bond is planar. The geometry was optimized on the excited state surface. Constraints on both bridge bonds were employed in the case of the imidazolol form, for which this geometry is unfavorable on the excited state potential surface. In the other cases, no constraints were used.

3. $(\theta_L, \theta_R) = \left(\frac{\pi}{2}, 0\right)$ : The geometry is twisted about the phenoxy-bridge bond, while the imidazoloxy-bridge bond is planar. Constraints on both bonds were used in the case of the phenol form, for which this geometry is unfavorable on the excited state potential surface. In the other cases, no constraints were used.

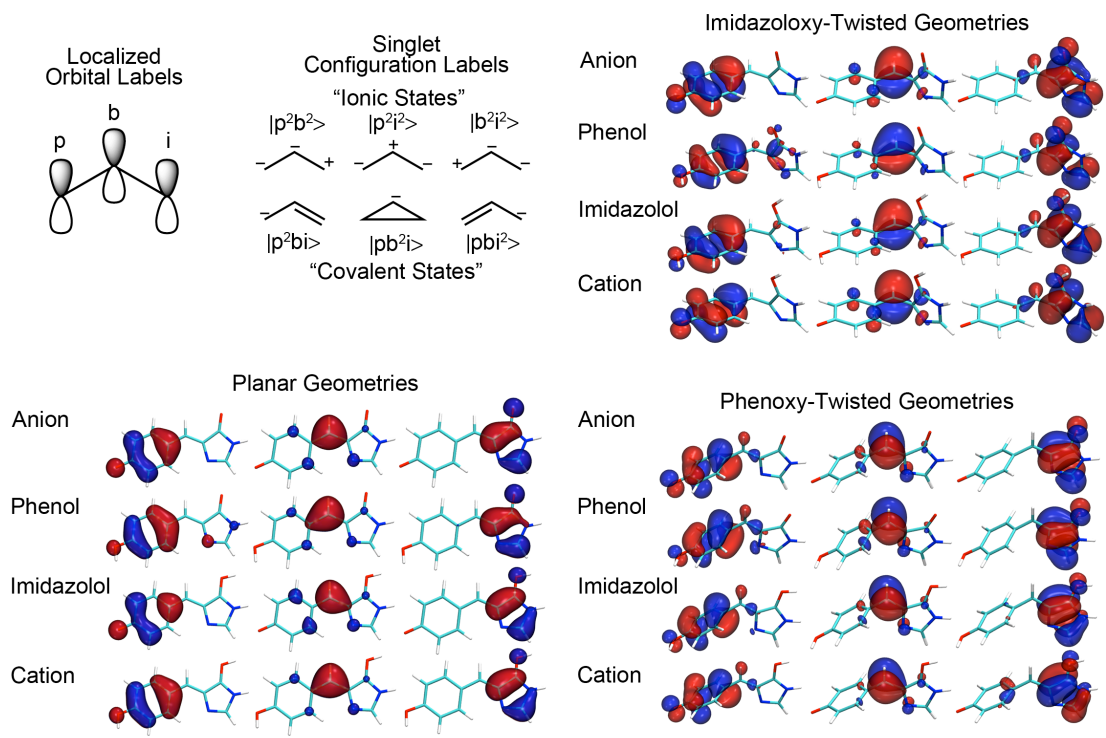

FIGURE 1. Boys-localized active space orbitals for GFP chromophore models at planar ground state (bottom left), imidazoloxy-twisted excited state (top right) and phenoxy-twisted excited state (bottom right) geometries. For each model and at each geometry, the orbitals are localized in chemically distinct regions corresponding to the rings and the bridge. With four electrons in the active space, these orbitals generate six singlet configurations for which valence-bond interpretations are possible (upper left). The analogous localization of the orbitals allows a common labeling scheme to be applied to the orbitals and configurations for all models (upper left).

Cartesian coordinates of all geometries used are listed in Tables S1-S12 of the Supplement, and internal coordinates (Z-matrices) used to construct the twisting channel models are available in Tables S13-S25 of the Supplement.

Previous results on an anionic GFP chromophore models have indicated that the low-energy pathways on the excited state correspond to displacement of a one torsion coordinate or the other[33, 35, 37, 38, 95-99]. Significant displacements along "mixed" torsion coordinates (e.g. hula-twist[100] displacements), are not favorable on the excited state surface[98]. These results suggest that the twisting channels corresponding to distinct bonds will give rise to distinguishable families of trajectories on the excited state surface, and it is justifiable to consider them independently.

Due to the asymmetry of the imidazoloxy ring about it's bond to the bridge, there are distinguishable isomers for these molecules where the imidazolol ring is oriented in a $Z$ or $E$ conformation with respect to the phenoxy-bridge bond[1]. We worked with the $Z$ isomers (i.e. those depicted in Scheme 1). The model (2.1) will not describe correctly the energetics of the complete $Z$-$E$ photoreaction if the isomers have different state energies. This could be fixed at the expense of adding additional term(s) with appropriate periodicity to the functional form describing the torsion[96], and would probably require at least one additional fit parameter. Results available in the literature suggest that the $Z$ and $E$ isomers have closely similar excitation energies and state-specific energies, and that the excited-state potential is similar for the two isomers near their ground-state geometries[33]. Ultrafast spectroscopy experiments on a model GFP chromophore in its $Z$ and $E$ isomeric states concluded that the decay times were indistinguishable, consistent with similar local excited-state potential surfaces[101]. We content ourselves with the goal of achieving a description of the local twisting channels in the region of the *Z* isomer only.

At each point along the interpolated twisting channel models, the SA2-CAS(4,3)*RS2//cc-pvdz adiabatic energies were obtained, as well as the SA2-CAS(4,3)//cc-pvdz dipole observables. There was no significant difference between the dipole observables obtained from the reference wavefunctions and the first-order perturbed wavefunctions.

Generalized Mulliken-Hush diabatic states for a two state system are obtainable directly by transformation of the Hamiltonian with the same transformation that diagonalizes the projection of the dipole operator along a chosen reference coordinate[62, 63]. This is equivalent to applying the Boys-Foster-Edminston-Ruedenberg[78, 79] localization algorithm to the many-electron adiabatic states in the two-state Hilbert space[62]. It would, therefore, also have been possible (and equivalent) to obtain the diabatic states (and associated parameters) directly from our quantum chemical calculations. We have, arbitrarily, chosen to fit against the adiabatic states instead.

The configuration space spanned by the localized active SA2-CAS(4,3) orbitals is analogous to the configuration space that has been used by Shaik, Hiberty and coworkers to construct valence-models of $S_N2$ reaction mechanisms[102-104]. An interesting alternative approach to the electronic structure would be to directly construct the valence-bond states implied in Schemes 1 and 2, using the strategies discussed in those works.

We note that only the anionic and phenol forms in Scheme 1 have been invoked in assignments of the spectra of GFP variants[105]. We invoke the imidazolol and cationic forms here because they allow sampling of complementary electronic structure at different proximity to the cyanine limit[48, 49], and are thus useful for testing the model in different regimes. There is a conserved hydrogen bond interaction between the imidazoloxy oxygen and a nearby arginine residue in all functional fluorescent proteins, and it may be appropriate to consider the imidazolol and oxonol cation forms as representing a strong limit of this interaction[106]. This notion is broadly supported by spectra of gas-phase chromophore models with positively charged amine groups coordinating the imidazoloxy oxygen, which absorb at energies intermediate between the phenol and cationic forms[107]. The oxonol cation form discussed here is distinct from the imine-protonated cationic form that has been discussed in the literature[105], and whose absorption has been recorded in solution[108].

In the context of the model (2.2), we will refer to the "cyanine limit" as the limit where $\delta \to 0$. This also implies that $\lambda(0,0) \to 0$, by Equation (2.2). This is in keeping with the empirical definition of the "cyanine limit" as a limit characterized by a well-known set of optical response properties of the equilibrium state (which, in our model, corresponds to the planar ground state).

## 3. Results

### 3.1. Model Parameterization

Table 1 lists the relevant energy and dipole parameters obtained from the adiabatic calculations at the planar and twisted geometries used to construct the twisting channel models. Table 2 lists the extracted Hamiltonian parameters for all protonation states of the GFP chromophore model highlighted in Scheme 1.

Table 1. Average energies (eV), adiabatic energy gaps (eV), difference dipole moments (eÅ), transition dipole moments (eÅ) and diabatic difference dipoles (eÅ, c.f. Equation (2.10)) obtained at each of the three geometries above for each of the four models shown in Scheme 1. The direction of projection of the dipole observables was taken as the largest principle component of the nuclear charge distribution.

| Model | $(\theta_L, \theta_R)$ | $\bar{E}$ | $\Delta E$ | $\Delta\mu$ | $\mu_{01}$ | $eR_{DA}$ |
|---|---|---|---|---|---|---|
| Anion | $(0,0)$ | 0.00 | 2.53 | -0.37 | -2.03 | 4.08 |
|  | $(0, \pi/2)$ | 0.56 | 0.56 | 3.04 | -0.03 | 3.04 |
|  | $(\pi/2, 0)$ | 0.40 | 1.11 | -3.37 | 0.00 | 3.37 |
| Cation | $(0,0)$ | 0.00 | 2.77 | -0.17 | 1.88 | 3.76 |
|  | $(0, \pi/2)$ | 0.55 | 1.00 | 2.96 | 0.06 | 2.96 |
|  | $(\pi/2, 0)$ | 0.40 | 1.32 | -3.08 | 0.01 | 3.08 |
| Phenol | $(0,0)$ | 0.00 | 3.56 | -1.17 | 1.54 | 3.29 |
|  | $(0, \pi/2)$ | 0.76 | 1.07 | -2.55 | 0.02 | 2.55 |
|  | $(\pi/2, 0)$ | 0.78 | 3.63 | -2.72 | -0.05 | 2.72 |
| Imidazolol | $(0,0)$ | 0.00 | 3.61 | 1.46 | -1.80 | 3.88 |
|  | $(0, \pi/2)$ | 0.46 | 3.00 | 2.65 | -0.26 | 2.70 |
|  | $(\pi/2, 0)$ | 0.53 | 1.11 | 3.33 | 0.00 | 3.33 |

**Table 2.** Model Hamiltonian parameters (eV, c.f. Equation (2.1)) of all of the (oxonol) protonation states of the GFP chromophore model shown in Scheme 1.

| Form | $\delta$ | $\varepsilon$ | $\gamma_R$ | $\gamma_L$ | $\gamma'_R$ | $\gamma'_L$ |
|---|---|---|---|---|---|---|
| Anion | -0.23 | 2.52 | 0.48 | 0.42 | 0.08 | -0.02 |
| Cation | -0.13 | 2.77 | 0.56 | 0.50 | -0.01 | -0.10 |
| Phenol | -1.27 | 3.33 | 0.43 | 0.98 | 0.33 | -0.20 |
| Imidazolol | 1.36 | 3.34 | 0.63 | 0.33 | -0.17 | 0.20 |

The parameter values suggest categorization according to the proximity to the cyanine limit. Previous work has shown that the anionic and cationic forms in Scheme 1 are close to the cyanine limit, while the phenol and imidazolol forms are far from the limit in opposite directions (i.e. $\lambda(0,0)$ has equal magnitude but opposite sign for these cases)[48, 49]. The forms close to the cyanine limit (anion and cation, c.f. Scheme 1) have approximately $\gamma_L \sim \gamma_R$ and $\gamma'_L \sim \gamma'_R$. If the dyes were symmetric, then this would be required. It is not here, but the approximate equality suggests that the effective potential felt by the frontier electrons is more symmetric than apparent from the nuclear skeleton. The "cross terms" $\gamma'$ are small for these forms. By this, we mean that they are smaller than the *a priori* expected accuracy (~±0.1-0.2eV) of the computational chemistry results to which the Hamiltonian was fit[109, 110].

For the species far from the cyanine limit (phenol and imidazolol, c.f. Scheme 1), there is a significant asymmetry in both the $\gamma$ and $\gamma'$ terms corresponding to the two bonds. This reflects the chemical notion that these forms possess significant polyenic character, with significant bond alternation. Also, for these forms the $\gamma'$ parameters have significant magnitude, in contrast to the cyanine-like species. The pattern seems to be that if the ground state is dominated one diabatic state (for example, $|L\rangle$, as for the phenol) then we have that $\gamma_R > \gamma_L$ and also $\gamma_L > 0 > \gamma_R$. This seemed counterintuitive to us at first, because the interpretation of the $\gamma$ parameters as "bond energies" seemed in conflict with the notion that the ground-state bond should carry a smaller penalty for twisting. The reason is that the parameters are not only for the ground state, but also for the excited state. They are best considered as characteristics of the state-averaged ensemble. In light of this, it is clear that if twisting a given bond is unstable in *both* states, the associated strain energy must be larger for that bond.

### 3.2. Model Validation

Figure 2 shows a comparison between the calculated SA2-CAS(4,3)*RS2//cc-pvdz energies along the model twisting channels. The comparison is good along the entire length of the channels in the excited state. The energies obtained from the parameterized model Hamiltonian (2.1) are within 0.1 eV of the calculated excited state energies in most cases, and in all cases are within 0.2eV of the calculated energies. For the points against which the model was fit, this is not a surprise. However, the Figure shows that the model is able to achieve a reasonable description of the *overall* shape of the potential surface characterizing the twisting channels. Estimates of the intrinsic accuracy of multireference perturbation theory based on complete active space wavefunctions are in the range of ±0.1-0.2eV[109, 110]. The model is able to describe the shape of the twisting channels to within this range.

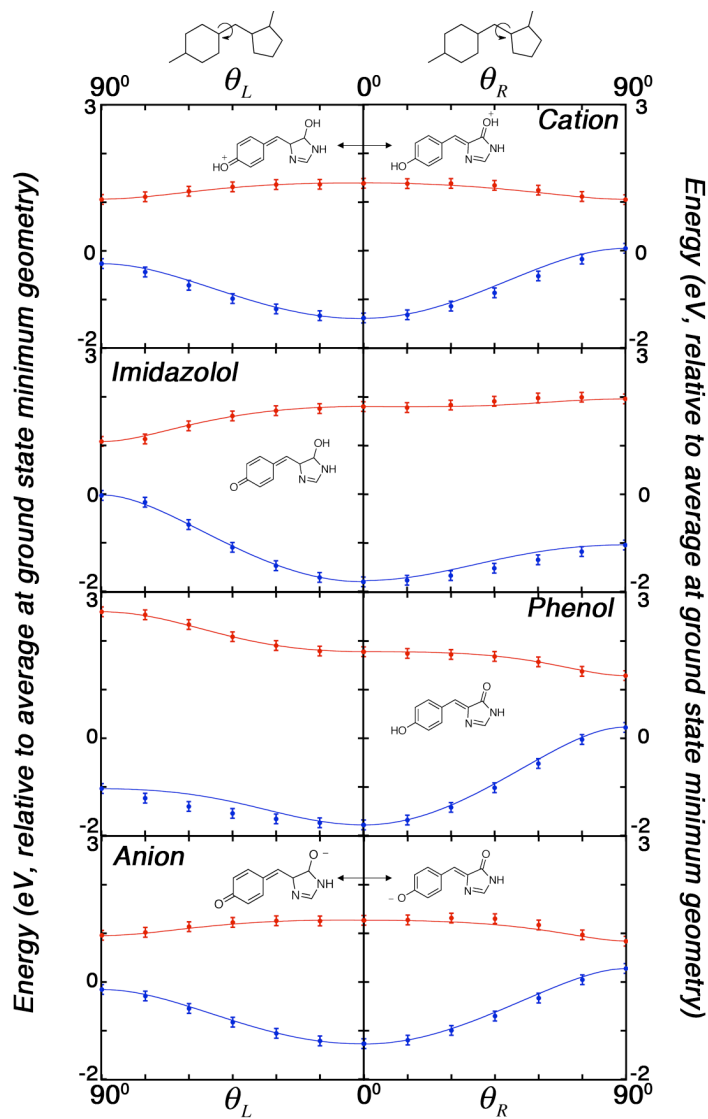

FIGURE 2. Energetics of twisting channels computed using the 2-state model (red and blue lines) against energies computed using SA2-CAS(4,3)*RS2//cc-pvdz (red and yellow circles) along synchronous transit coordinates connecting the ground state minimum with optimized twisted structures optimized on the excited state. Optimizations are described in the text; synchronous transit paths are linear interpolations in a set of internal coordinates. Error bars on the computational data are set at ±0.1eV and are meant to aid in quantitative appraisal of the fit. The zero of the energy scale is set to the average energy at the ground state minimum geometry.

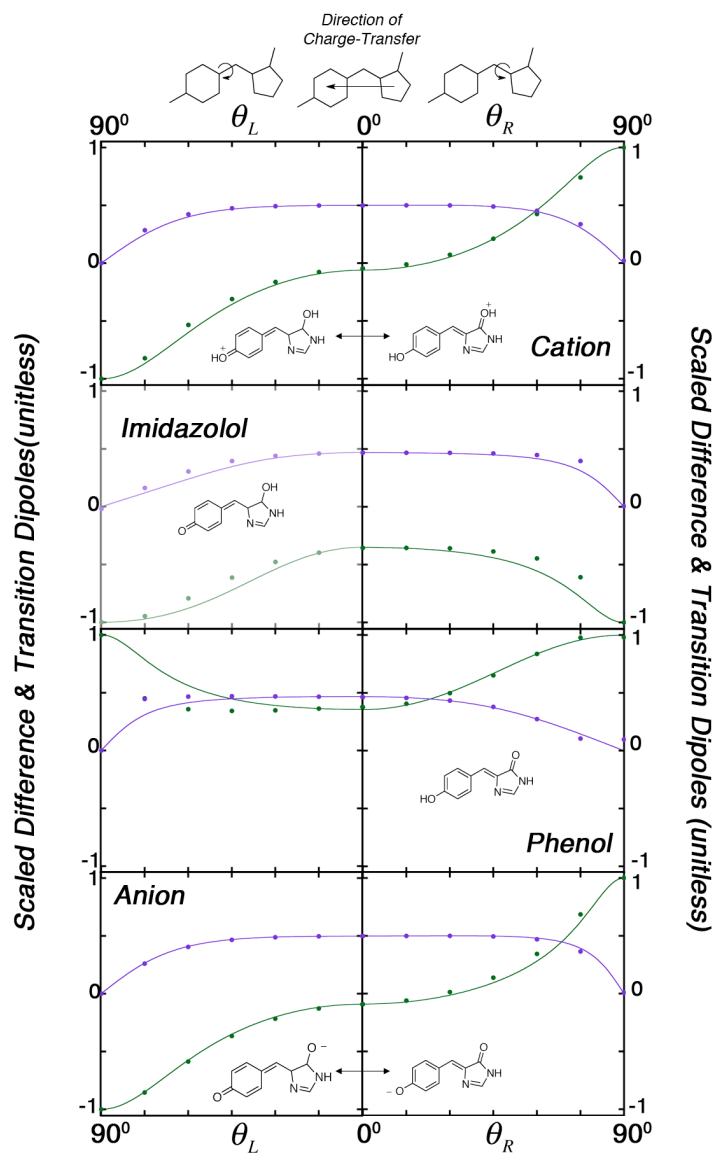

FIGURE 3. Dimensionless transition (violet) and difference (green) dipole observables for the lowest-lying electronic transition of the models in Scheme 1, evaluated using the 2-state Hamiltonian model (solid lines) and by SA2-CAS(4,3)//cc-pvdz calculations (dots). The dimensionless observables are obtained by scaling with the Generalized Mulliken-Hush donor-acceptor dipole (Equation (2.10)). The observables were evaluated along synchronous transit coordinates connecting the ground state minimum to optimized excited state structures twisted about the distinct bridge bonds.

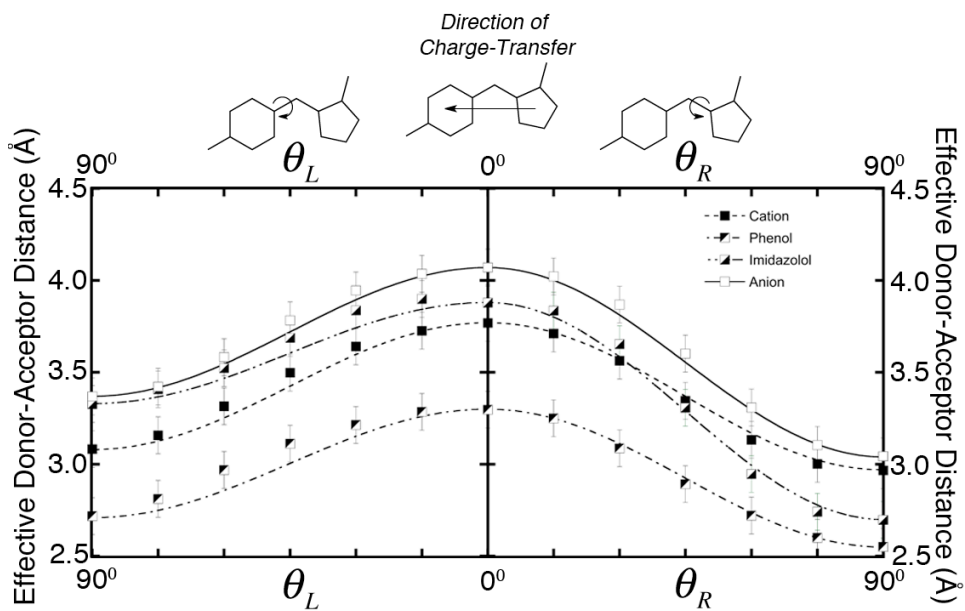

Figure 4. Generalized Mulliken-Hush donor-acceptor distance evaluated along the twisting channels. The effective distance $R_{DA}$ was evaluated using adiabatic dipole observables along synchronous transit pathways connecting the ground state geometry with either of two optimized excited state geometries twisted about two bridge bonds. Error bars are set at ±0.1 Å and are meant to assist in quantitative appraisal of the plot. Lines represent the functional form indicated in the text (c.f. Equation (3.1)).

Figure 3 shows the values of the dimensionless difference dipole $\Delta\tilde{\mu}(\theta_L, \theta_R)$ and the dimensionless transition dipole $\tilde{\mu}_{01}(\theta_L, \theta_R)$ along the twisting channels. Again, a good description is obtained along the entire length of the channels. In particular, the model captures the qualitative differences between the resonant (anion, cation) and non-resonant (phenol, imidazolol) forms. It correctly describes the twist-dependent reversal of polarization for the cases near the cyanine limit, and the absence of this reversal in the cases far from the limit.

The dimensionless dipole observables shown in Figure 3 are scaled by the Generalized Mulliken-Hush diabatic difference dipole[63] $eR_{DA}$ (c.f. Equation (2.10)), which varies with the molecular geometry. In order to quantify the actual field generated by the polarization in the excited state, it is useful to assess the magnitude of $eR_{DA}$ as a function of progress along the twisting channels. This data is collected and displayed in Figure 4 for all of the forms in Scheme 1. We find that the variation along the twisting channels can be described to good accuracy by the interpolation

$$R_{DA}(\theta_L, \theta_R) = R_{DA}(0,0)\cos^2\theta_L \cos^2\theta_R + R_{DA}\left(0, \frac{\pi}{2}\right)\sin^2\theta_R + R_{DA}\left(\frac{\pi}{2}, 0\right)\sin^2\theta_L \qquad (3.1)$$

Figure 4 shows that the variation of the effective dipole length is modest. For any of the protonation states studied, the variation of the effective distance over both twisting channels amounts to ~10% of its mean value, and the spread over all protonation states at any particular twist coordinate shown is also ~10% of the mean. The magnitude of the effective distance is of the appropriate magnitude to describe transfer of a charge between frontier states localized on separate rings but delocalized within each ring domain (c.f. Figure 1). This suggests that, in future applications of the model, it would not be unreasonable to approximate $eR_{DA}$ as constant.

We note that $R_{DA}$ does not correlate quantitatively with the geometric distances such as e.g. the distance between the oxygens bearing the formal charge in Schemes 1 and 2, or the width of the methine bridge. It is best understood as a mean distance characterizing the transfer between fragment orbitals localized on the rings (c.f. Figure 1). Variations in the value of $R_{DA}$ that do not reflect changes in the nuclear frame can be taken to reflect changes in the precise shape of the orbitals between which the charge is transferred (c.f. Figure 1).

The reason that the twist-dependence of the effective charge-transfer distance (3.1) has a similar $\sin^2\theta$ dependence as the diabatic energies (c.f. Equation (2.1)) is because both the energy and the dipole moments of the diabatic states are modulated by their projection on the "ionic" configurations associated with the bridge bonds. These projections are described within the four-electron, three-orbital model of the electronic structure that we use[49, 98, 111]. The bonding energy itself is a consequence of mixing between covalent and ionic states on the bonds[112, 113] (c.f. Figure 1). The coupling between these configurations is eliminated by twisting the bridge bonds, resulting in their removal from the configuration expansion of the diabatic states at the twisted geometries[98]. The removal of the ionic states manifests itself in a rise of the relevant diabatic state energies *and a reduction in the magnitude of their dipole moments*.

### 3.4. Twist-Coupled Charge Transfer and Charge-Transfer Intersections

The charge-transfer character of the excitation is conveniently quantified by the dimensionless difference dipole $\Delta\tilde{\mu}(\theta_L,\theta_R) \in (-1,1)$, which denotes the asymmetry of mixing of the diabatic states $|L\rangle$ and $|R\rangle$ into the adiabatic states $|0\rangle$ and $|1\rangle$. When $\Delta\tilde{\mu}(\theta_L,\theta_R) \to -1$, we have $|0\rangle \to |L\rangle$ and $|1\rangle \to |R\rangle$ and conversely for $\Delta\tilde{\mu}(\theta_L,\theta_R) \to 1$. The sign of $\Delta\tilde{\mu}(\theta_L,\theta_R)$ passes through zero on the contour defined by

$$\delta + 2(\gamma_R - \gamma'_R)\sin^2\theta_R - 2(\gamma_L - \gamma'_L)\sin^2\theta_L = 0 \ . \tag{3.2}$$

Equation (3.2) is the condition that must be satisfied for the diabatic energy splitting to vanish. The condition for a conical intersection in the model is obtained when the diabatic splitting and the coupling vanish simultaneously[114]. Setting the diabatic gap to zero yields the condition (3.2), while setting the coupling to zero yields

$$\varepsilon \cos\theta_L \cos\theta_R = 0 \tag{3.3}$$

Both conditions (3.2) and (3.3) must hold for a conical intersection to occur in the model (2.1).

Equation (3.2) implies that a reversal of the charge-transfer polarity of the excitation is *necessary* for the occurrence of a conical intersection in the model. Moreover, because equation (3.3) will always be satisfied on the lines given by $\theta_L = \frac{\pi}{2}$ and $\theta_R = \frac{\pi}{2}$, a reversal of the polarity is also *sufficient* for the occurrence of a conical intersection.

Equation (3.2) can be solved if

$$-2(\gamma_R - \gamma'_R) \leq \delta \leq 2(\gamma_L - \gamma'_L) \tag{3.4}$$

A typical situation for which Equation (3.4) is true is for the parameter set characterizing the anionic form of the GFP chromophore model (c.f. Table 1). This situation is illustrated in Figure 5, which shows the excitation charge-transfer polarity using a color scale. The contour line depicting the condition (3.2) is shown by a white dashed line, and separates regions of opposing charge-transfer polarity. When it intersects the line $\theta_R = \frac{\pi}{2}$, Equation (3.3) is also fulfilled and a conical intersection is found to occur at that point. Moreover, the Figure shows that although the change in polarity occurs gradually at near-planar geometries, it becomes more abrupt as the intersection is approached, and becomes arbitrarily sharp at the location of the intersection, at which point the derivative of (2.8) diverges across the contour.

Twisted conical intersections have been shown to occur on the potential surfaces of anionic GFP chromophore models[33]. We have verified that the minimal-energy intersections for the cationic form are qualitatively identical to those of the anion, indicating again their very similar electronic structure. A common feature is significant pyramidalization of the methine carbon[33]. Depending on the exact definition of the torsion coordinates, this pyramidalization can appear as a mixed twisting of the bonds[33]. In general, the intersections described by Equations (3.2) and (3.3) will occur when one bond is completely twisted and the other partially twisted; generally, the degree of total twist is somewhat greater than observed in the quantum chemical calculations[33].

The condition (3.4) is not fulfilled for the parameters characterizing the phenol and imidazolol GFP chromophore models (c.f. Table 2). Quantum chemical calculations indicate that a low-lying twisted conical intersection seam does exist in the phenol form[38, 95], and we have verified that analogous intersections occur for the imidazolol form (although the twist distribution on the bridge is reversed). The geometry of the minimal energy point on the intersection seam for the neutral form is characterized by significant puckering of the imidazolol ring[38, 95]. Our model has no degree of freedom that can plausibly represent the pyramidalization of the bridge-adjoining sites on the rings. Describing these low-energy intersections for species far from the cyanine limit may be possible in an extended model; we leave this for future work.

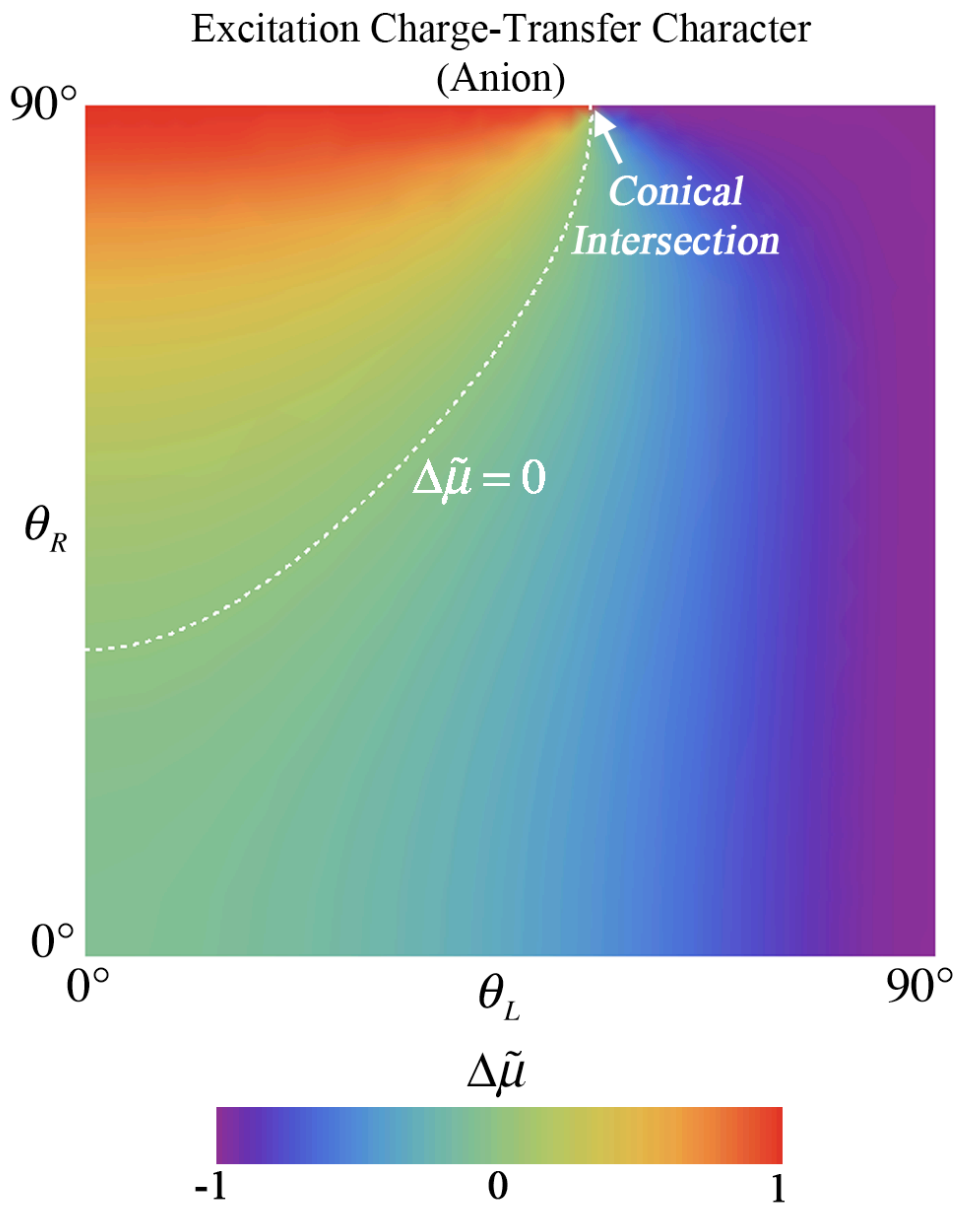

FIGURE 5. Twist-dependent charge-transfer polarity of the excitation described in the model (2.1) for parameters appropriate to the anionic GFP chromophore (c.f. Table 2, first row). The charge-transfer character is shown using the dimensionless adiabatic difference dipole defined in Equation (2.8) (color scale at bottom). The charge-transfer character vanishes on the dashed white line. At the point where this line intersects the edge of the plot corresponding to $\theta_R = \frac{\pi}{2}$, a conical intersection occurs, and the change in the charge transfer becomes arbitrarily sharp across the contour.

**3.5. Excited-state driving forces and channel-specific excited state biasing**

The model Hamiltonian (2.1) can describe the occurrence of dual low-energy twisting channels on the excited state surface for systems close to the cyanine limit, as well as the selection of individual low-energy channels far from the cyanine limit (c.f. Figure 2).

In a viscous medium, one expects that the timescale for displacement along a twisting channel would be given by equating the force (actually, torque) $F = -dE/d\theta$ to the opposing viscous force leading to[115]

$$\tau = \frac{2\eta V}{F}, \qquad (3.5)$$

where $V$ is the volume of the rotor (here, the phenoxy or imidazoloxy rings) and $\eta$ is the medium viscosity[115]. For $\eta \sim 1\text{cP}$, $V \sim 30$ Å$^3$, $F \sim 0.1$ eV/rad, this implies $\tau \sim 1$ ps, which is the right order of magnitude for many chromophores. The relation above implies that the lifetime scales linearly with the viscosity, which has been observed over four orders of magnitude for thioflavin-T[115]. However, many chromophores, including GFP chromophore models[20, 116, 117], exhibit non-radiative lifetimes with sub-linear fractional power law dependence on viscosity, $\eta^\alpha$[118]. For GFP models, the value of $\alpha$ has been estimated at ~0.5[116, 117], but other work indicated a lower value of ~0.25[20]. Some results indicate that different timescales contributing to the (generally, multi-exponential decay) may have different viscosity scaling[25].

In a simple approximation, we can consider the *mean* driving forces $\bar{F}$ for the excited state twisting channels,

$$-\frac{\pi}{2}\bar{F}_L = \Delta_L E_1 = E_1\left(\frac{\pi}{2}, 0\right) - E_1(0,0)$$
$$-\frac{\pi}{2}\bar{F}_R = \Delta_R E_1 = E_1\left(0, \frac{\pi}{2}\right) - E_1(0,0)$$
(3.6)

where $\Delta_{X \in \{L,R\}} E_1$ is the *channel bias*, equal to the excited-state energy of the twisted geometry associated with the channel, relative that of the planar geometry. Figure 5 illustrates this with a schematic. The channel biases can be written as (c.f. Equations (2.4) – (2.6))

$$\Delta_L E_1 = (\gamma_L + \gamma'_L) + \frac{1}{2}|\delta - 2(\gamma_L - \gamma'_L)| - \frac{1}{2}\sqrt{\varepsilon^2 + \delta^2}$$
$$\Delta_R E_1 = (\gamma_R + \gamma'_R) + \frac{1}{2}|\delta + 2(\gamma_R - \gamma'_R)| - \frac{1}{2}\sqrt{\varepsilon^2 + \delta^2}$$
(3.7)

In the limit of a symmetric monomethine dye where $\gamma_L = \gamma_R = \gamma$, $\gamma'_L = \gamma'_R = \gamma'$, $\delta = 0$ the twisted states have the same energy and (3.7) reduces to

$$\Delta_L E_1 = \Delta_R E_1 = 2\gamma - \frac{\varepsilon}{2}$$
(3.8)

where we have assumed that $\gamma > |\gamma'| \geq 0$ as is the case for the systems described in Table 2 (and which appears also to be true for other monomethine systems we have examined so far), and also that $\varepsilon > 0$ (which we can do without loss of generality for the Hamiltonian (2.1)).

We say a channel $X \in \{L, R\}$ is "open" if the associated channel bias is negative ($\Delta_{X \in (L,R)} E_1 \leq 0$). For a symmetric monomethine dye, (3.8) indicates that both twisting channels will be open if

$$\frac{4\gamma}{\varepsilon} \leq 1$$
(3.9)

For a dye that is at the cyanine limit (i.e. "resonant"[119]) but not necessarily symmetric (i.e. $\delta = 0$, but $\gamma_L \neq \gamma_R$, $\gamma'_L \neq \gamma'_R$. Then the channel biases are different

$$\Delta_L E_1 = 2\gamma_L - \frac{\varepsilon}{2}$$
$$\Delta_R E_1 = 2\gamma_R - \frac{\varepsilon}{2}$$
(3.10)

and the conditions for the channels to be open are, respectively,

$$\frac{4\gamma_L}{\varepsilon} \leq 1$$
$$\frac{4\gamma_R}{\varepsilon} \leq 1$$
(3.11)

Equations (3.10) and (3.11) are only slight generalizations of Equations (3.8) and (3.9), respectively.

Equations (3.9) and (3.11) show that the accessibility of the excited state twisting channels is determined by competition between the interstate coupling and the energy penalties associated with bond twisting.

For the general case, we find that there is a derivative discontinuity in the twisting channel biases $\Delta_{X \in \{L,R\}} E_1$ with respect to changes in $\delta$, which occurs when the argument of the absolute value in (3.7) changes sign. The bias for the $L$ channel on either side of the derivative discontinuity is

$$\delta > 2(\gamma_L - \gamma'_L) \rightarrow \Delta_L E_1 = 2\gamma'_L + \frac{\delta}{2} - \frac{\varepsilon}{2}\sqrt{1 + \left(\frac{\delta}{\varepsilon}\right)^2}$$
$$2(\gamma_L - \gamma'_L) > \delta \rightarrow \Delta_L E_1 = 2\gamma_L - \frac{\delta}{2} - \frac{\varepsilon}{2}\sqrt{1 + \left(\frac{\delta}{\varepsilon}\right)^2}$$
(3.12)

The conditions for the $L$ channel to be open are then

$$\delta > 2(\gamma_L - \gamma'_L) \rightarrow \frac{4\gamma'_L}{\varepsilon} \leq \sqrt{1 + \left(\frac{\delta}{\varepsilon}\right)^2} - \frac{\delta}{\varepsilon}$$
$$2(\gamma_L - \gamma'_L) > \delta \rightarrow \frac{4\gamma_L}{\varepsilon} \leq \sqrt{1 + \left(\frac{\delta}{\varepsilon}\right)^2} + \frac{\delta}{\varepsilon}$$
(3.13)

For the *R* channel we have

$$-2(\gamma_R - \gamma'_R) > \delta \rightarrow \Delta_R E_1 = 2\gamma'_R - \frac{\delta}{2} - \frac{\varepsilon}{2}\sqrt{1 + \left(\frac{\delta}{\varepsilon}\right)^2}$$
$$\delta > -2(\gamma_R - \gamma'_R) \rightarrow \Delta_R E_1 = 2\gamma_R + \frac{\delta}{2} - \frac{\varepsilon}{2}\sqrt{1 + \left(\frac{\delta}{\varepsilon}\right)^2}$$
(3.14)

and the conditions for the channel to be open are

$$\delta > -2(\gamma_R - \gamma'_R) \rightarrow \frac{4\gamma_R}{\varepsilon} \leq \sqrt{1 + \left(\frac{\delta}{\varepsilon}\right)^2} - \frac{\delta}{\varepsilon}$$
$$-2(\gamma_R - \gamma'_R) > \delta \rightarrow \frac{4\gamma'_R}{\varepsilon} \leq \sqrt{1 + \left(\frac{\delta}{\varepsilon}\right)^2} + \frac{\delta}{\varepsilon}$$
(3.15)

The conditions (3.13) and (3.15) reduce to the conditions for a symmetric or resonant dye under the appropriate restrictions on the parameter values. No dimensions are carried on either side of Equations (3.13) and (3.15), so the availability of the excited state twisting channels is independent of the overall energy scale.

The twisting channel biases (Equations (3.12) and (3.14)) have opposing dependence on $\delta$ for small values of $-2(\gamma_R - \gamma'_R) < \delta < 2(\gamma_L - \gamma'_L)$, but outside of this region the dependence on $\delta$ will be similar for both channels. This point is illustrated in Figure 7, which shows the dependence of the channel bias on $\delta$.

The derivative discontinuities on the boundary of the region $-2(\gamma_R - \gamma'_R) \leq \delta \leq 2(\gamma_L - \gamma'_L)$ are intimately connected to the behavior of conical intersections in the model. It is at these points that the endpoints of the twisting channels coincide with the conical intersection. The discontinuity in the derivative of the excited state potential carries over into the driving forces through their definitions (3.6)-(3.7). These points also bound the region of parameter space where conical intersections can occur in the model (c.f. Equation (3.4)).

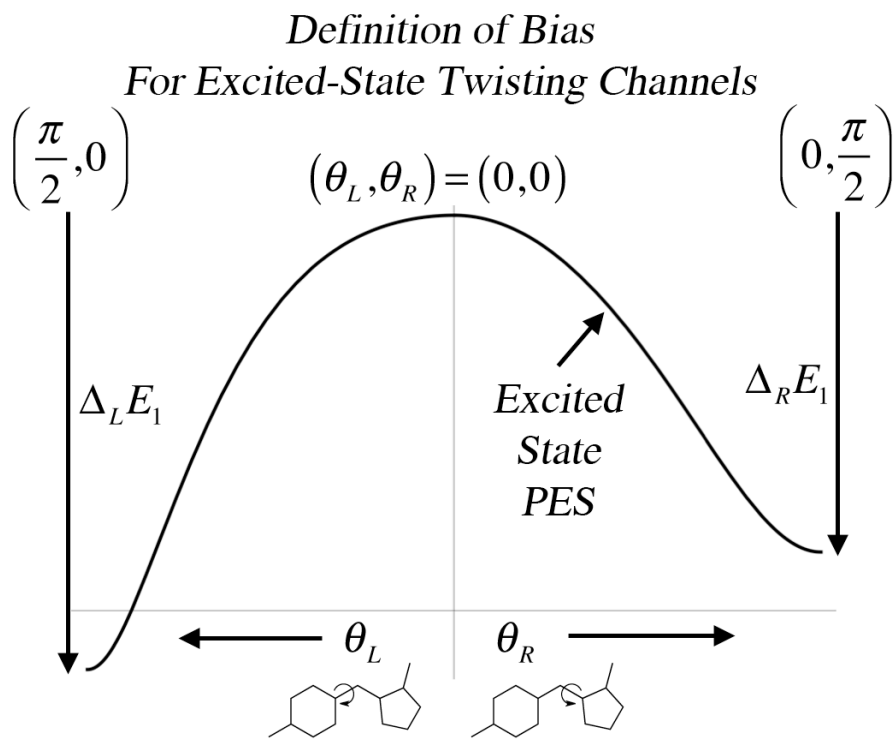

FIGURE 6. Schematic depiction of excited-state twisting channel driving forces (defined in (3.5)). The driving "forces", which have units of energy (or torque) are literally integrated forces. They are defined as the excited-state potential energy of the twisted structures representing the distinct channels, relative to that at the planar ground-state structure.

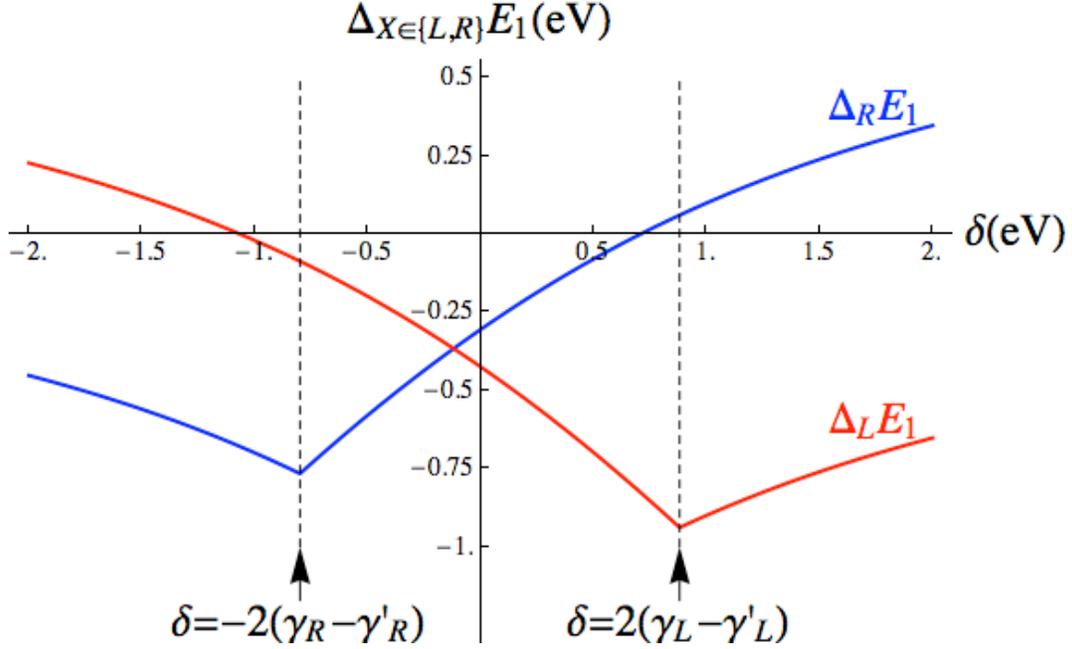

FIGURE 7. Excited-state twisting channel biases (c.f. Equation(3.7)) for different values of the diabatic biasing potential $\delta$ (c.f. Equation (2.1)). Parameters (excepting $\delta$) were taken as appropriate to the anionic state of the GFP chromophore (c.f. Table 2, top row). Driving forces $\Delta_L E_1$ and $\Delta_R E_1$ for the $L$ and $R$ channels are shown in red and blue, respectively. Dashed lines demarcate the regions of qualitatively different behavior; between these points the dependences of the driving forces on $\delta$ are opposite. Outside this region the driving forces have identical dependence on $\delta$, and only differ by a constant. Derivative discontinuities at the dashed lines indicate the points at which the twisting channel endpoint coincides with a conical intersection.

One point that is highlighted by Figure 7 is that, in general, the driving forces for the different channels will not be equal at the cyanine limit (defined as $\delta = 0$). At the limit, the difference in the driving forces will be

$$\Delta_R E_1 - \Delta_L E_1 \big|_{\delta=0} = 2(\gamma_R - \gamma_L) \ . \tag{3.16}$$

We can also solve for the value of $\delta$ where the driving forces for the different channels are equal. By the arguments above, this point must lie in the region $-2(\gamma_R - \gamma'_R) \leq \delta \leq 2(\gamma_L - \gamma'_L)$. By setting Equations (3.12) and (3.14) equal we obtain

$$\delta\big|_{\Delta_R E_1 = \Delta_L E_1} = 2(\gamma_L - \gamma_R) \ . \tag{3.17}$$

Outside the region $-2(\gamma_R - \gamma'_R) \leq \delta \leq 2(\gamma_L - \gamma'_L)$, the driving forces differ by a constant. Again, using Equations (3.12) and (3.14) we obtain

$$\begin{aligned}\Delta_R E_1 - \Delta_L E_1 \big|_{2(\gamma_L - \gamma'_L) < \delta} &= 2(\gamma_R - \gamma'_L) \\ \Delta_R E_1 - \Delta_L E_1 \big|_{\delta < -2(\gamma_R - \gamma'_R)} &= 2(\gamma'_R - \gamma_L)\end{aligned} \ . \tag{3.18}$$

Equations (3.16) - (3.18) show that if information about the driving forces can be obtained, then this can be used to infer the magnitude of the relevant $\gamma$ terms.

### 4. Discussion

#### 4.1. General discussion

We have described a two-state model Hamiltonian that describes the coupling of intramolecular charge-transfer behavior to excited state twisting in monomethine dyes near and far from the cyanine limit. The model is based on a generalized Mulliken-Hush two-state model[62, 63] where the energies and couplings of the diabatic[68] states are coupled to the twisting motions using valence-bond-inspired[69-72] functional dependencies. The model is capable of describing the potential energies along the excited-state twisting channels (c.f. Figure 2), and the evolution of charge-transfer behavior that accompanies twisting (c.f. Figure 3). The model captures the twist-dependence of the charge-transfer polarization in dyes close to cyanine limit, as well as the disappearance of this dependence away from the cyanine limit. This distinction between near-resonance and far-from-resonance regimes has been previously documented in studies of the green fluorescent protein chromophore (a model methine dye)[50]. This paper also extends these results, by showing that complementary protonation states, which access different regimes of detuning from the cyanine limit, show complementary behavior. The model is found in each of these cases to describe the coupling of the charge transfer to progress along the twisting channels.

We have parameterized and evaluated the model on several (oxonol) protonation states of a model of the green fluorescent protein[10-12, 14] (GFP) chromophore. This particular example is of general interest to photochemical physics because of the wide range of photobehavior that can be elicited in different environments[1, 16, 105, 120]. One behavior of particular interest is the control of the fluorescence quantum yield of the chromophore, which can be modulated over several orders of magnitude in different environments[1, 105]. The modulation is related to the suppression of competing excited-state twisting processes that lead to radiationless decay[1, 105]. Although only the phenol and anion models in Scheme 1 have been invoked in assignments of the spectrum of GFP variants[105], there is a hydrogen-bonding interaction with a conserved arginine residue for which the oxonol cation and imidazolol form may be considered as a strong-interaction limit[106]. Other forms are not considered here but, based on our previous results[48], we infer that our model can also describe twisting processes in these forms.

The model that we describe can be elaborated into a strategy for introducing twist-dependent polarization into classical force-field descriptions of monomethines. The excited state twisting dynamics of a GFP chromophore model has recently been studied[96] in protein and vacuum using molecular dynamics force fields that could describe the twisting channel bifurcation, but did not describe the twist-dependent polarization. The current paper offers a simple strategy to couple the charge distribution moments of monomethine dyes to the twisting modes; it is conceivable that the same strategy might be extended to confer twist-dependent charge distributions to molecular mechanics models of fluorescent protein chromophores and other biotechnologically important monomethines.

The binding-dependent fluorescence enhancement due to suppression of excited-state twisting processes is not specific to the GFP chromophore, but is apparently a general characteristic of monomethine dyes (i.e. di- and triarylmethanes and monomethine cyanines)[3, 4, 6, 121].

The radiationless decay of monomethine dyes in condensed phase has been studied as an example of a barrierless viscosity-controlled process[21, 28, 39-41]. Predictions of internal charge-transfer coupled to the twisting motion have been known for some time[26, 31]. Theoretical models that have been used to interpret experiments describe the reaction as an overdamped motion on either a parabolic or flat potential surface with position-dependent sinks[42, 64]. These models do not allow a separation of rate contributions from mechanical friction and dielectric friction or solvent reorganization effects[122]. Moreover, they do not describe bifurcation of the twisting channels, such as predicted for dyes close to the cyanine limit[50]. The model we describe here, of the potential surfaces and charge-transfer behavior, can be used as input in dissipative dynamics simulations, which may treat separately the effects arising from mechanical friction and dielectric relaxation effects.

The apparent dependence of the twisting channel on the protonation state of fluorescent protein chromophore models appears also to hold for the chromophores of photoactive yellow protein[123] (which have also an oxonol electronic structure). In the latter case, simulations have suggested that protonation and field effects can influence the driving forces for distinct twisting channels[124-127]. Experimental studies indicate that titration affects the solvent-dependence of the decay rate[128]. The possibility of charge-transfer deactivation pathways has been noted[129]. Proton transfer and excited state twisting reactions are closely linked in the protein photocycle[130]; a similar coupling between these processes has been noted in certain photoswitchable fluorescent proteins[17].

The model describes an intimate physical relationship between charge-transfer behavior and the shape of the excited state potential along the twisting channels. We find that twist-dependent polarization switching only occurs for parameter regimes where a conical intersection can occur, and that the intersection lies along the manifold of geometries where the adiabatic difference dipole vanishes. Outside of this parameter range we find that the magnitude of the difference dipole may change but its direction will not. One may expect that this behavior will be relevant to the excited state twisting dynamics of the model upon coupling with a responsive polarizable environment. Such investigations are currently being pursued in our group.

The model predicts that differential selection between the twisting channels by an applied diabatic potential can only be achieved in a limited region around the cyanine limit. This region coincides with the region where conical intersections and polarization reversal can occur. The parameter region is bounded by derivative discontinuities in the excited state channel biases, which are a consequence of the conical intersections. The extent of this region is determined by the relative magnitude of the diabatic biasing potential and the "exchange" energies associated with bond twisting. Within the region, the channel driving forces have opposing dependence on the diabatic biasing potential. Outside of it, they have the same dependence and so differential channel selection can no longer occur. This feature of the model is of interest because our choice of a generalized Mulliken-Hush[62, 63] diabatic representation implies that the linear response to a homogeneous electric field can be described by via changes in the diabatic biasing potential. The implications of this to understanding channel selection in proteins and solution will be elaborated in future work.

**Conclusion**

We have described a two-state model Hamiltonian that can describe the changes in the potential surface and twist-dependent polarization of monomethine dyes near and far from the cyanine limit. We have parameterized the model against multireference perturbation theory calculations of the ground and excited state of four oxonol protonation states of the green fluorescent protein chromophore, which sample electronic structures at different proximities to the cyanine limit. The model can describe the qualitatively different relationships between twisting and charge transfer in each case. We have shown that preferential selection of one twisting channel over another can be accomplished by the application of a diabatic biasing potential only in a limited region of parameter space near the cyanine limit. We have derived quantitative expressions for the dependence of the driving forces associated with the twisting channels, and conditions for the channels to be open and closed. These expressions show that the availability of distinct twisting channels in a monomethine dye depends on the relative balance of the diabatic biasing potential to the "exchange" terms that describe the bond twisting energetics.

ASSOCIATED CONTENT

**Supplement** Cartesian coordinates (Å), state-specific SA-CASSCF energies, state-averaged SA-CASSCF natural orbitals and occupation numbers, MS-RSPT2 energies, and Z-matrices used to generate twisting channel models are available upon request in a Supplement.

AUTHOR INFORMATION

**Seth Olsen**

*School of Mathematics and Physics, The University of Queensland, Brisbane QLD 4072 Australia

**Author Contributions**

The manuscript was written through contributions of all authors. All authors have given approval to the final version of the manuscript.

**Funding Sources**

ARC Discovery Projects DP110101580 and DP0877875.

**Notes**

Any additional relevant notes should be placed here.

**ACKNOWLEDGMENT**

This work was funded by ARC Discovery Projects DP110101580 and DP0877875. All computations were carried out at the National Computational Infrastructure Facility under the auspices of Merit Allocation Scheme Grants m03 and n62. We thank A. Painelli, J. Reimers,

[122] It can be argued that the Ben-Amotz-Harris/Oster-Nishijima model does describe this, through the placement of two of sinks at different distances from the initial configuration. However, it is not clear that the authors considered this interpretation of this feature of the model.